\documentclass[showpacs,preprintnumbers,amsmath,amssymb]{revtex4}

\usepackage{graphicx}
\usepackage{dcolumn}
\usepackage{bm}

\begin{document}


\title{Wandering in Color-Space\footnote{To appear in  {\em Teller Memorial Volume of Acta Physica
Hungarica A (Heavy Ion Physics), 2004}} \\
 -- why the life of pentaquark is so long ? --}

\author{Yuu Maezawa$^1$}
 \email{maezawa@nt.phys.s.u-tokyo.ac.jp}
\author{Toshiki Maruyama$^2$}
 \email{maru@hadron02.tokai.jaeri.go.jp}
\author{Naoyuki Itagaki$^1$}
 \email{itagaki@phys.s.u-tokyo.ac.jp}
\author{Tetsuo Hatsuda$^1$}
 \email{hatsuda@phys.s.u-tokyo.ac.jp}
 \affiliation{
$^1$Department of Physics, University of Tokyo, Tokyo 113-0033, Japan \\
$^2$Advanced Science Research Center, Japan Atomic Energy Research Institute,
  Tokai, Ibaraki 319-1195, Japan
}
%

\begin{abstract}
The problem of the long life time of the pentaquark $\Theta^+$
 is investigated on the basis of the color molecular dynamics simulation.
 We find that it takes a long time (typically of $50-100$ fm/$c$) 
  for the initial pentaquark-state to 
 rearrange its color and spatial positions to decay into the nucleon + kaon
  final state. Structure of the 
  potential surface in the color and position spaces also supports
  this picture. Pentaquark wanders on the potential surface to find
   a narrow channel to decay.
\end{abstract}

\maketitle

\section{Introduction}

An exotic baryon, pentaquark $\Theta^+$, has been first reported
 by LEPS collaboration at SPring-8 \cite{exp1}. Later,
 there are other experiments reporting the confirmation of its
  existence, which is  summarized in refs. \cite{nakano04,PDG04}.
  Since the $\Theta^+$ decays into a neutron ($n$), and a kaon ($K^+$),
  its quark content is considered to be $uudd\bar{s}$.
 The mass is found to be about $1540\, \mathrm{MeV}$, while 
   there is yet no direct measurement of its spin and isospin,
 although $I=J=1/2$ is rather plausible. Its parity is also not
  known.  Its width ($< 10$ MeV)
   is exceptionally narrow as for a hadron resonance
 located at
  $110\, \mathrm{MeV}$ above the $nK^+$ threshold.
 There have been many theoretical analyses on the pentaquark 
 before and after the discovery of $\Theta^+$
  on the basis of the skyrme model, quark models, QCD sum rules,
   lattice QCD simulations and so on, which are
  summarized in  recent reviews \cite{oka04,sasaki04}.

 The question we address in this paper is the origin of the narrow
  width of the pentaquark which is the most peculiar feature of this
   new resonance. 
  Since there is no known selection rule from symmetry 
  to make the width naturally small,
  the narrow width should have some dynamical origin.
  In the past, there have been several attempts to
  explain the narrow width from combinatorial 
   suppression from the  spin-flavor  and color-factors 
  or from the special spatial structure due to diquark
   correlations.  However, no definite conclusion has been reached yet.
 
 In this paper, we propose a possible dynamical mechanism 
 for the long life time of the pentaquark by treating its
 hadronic decay  on the basis of the constituent quark model
  and molecular dynamics.
 Suppose that the colors of the constituent quarks 
 are ``well-mixed" inside the pentaquark (the 
  precise meaning of the mixing will be given later), then it takes a long time
  for the quarks to rearrange their colors, flavors, spins, 
  and spatial positions
  into two color-white clusters, i.e. the nucleon $N$ and the kaon $K$.
  The dynamical time scale of this rearrangement is 
 governed by the strong interactions among quarks and is
  not simply related to the ``distance" in color-flavor-spin-space
  between the pentaquark and the $NK$ state.
  Therefore,
  it is not trivial from the outset that the life time  is long. 

  The color molecular
  dynamics (CMD) simulations originally developed in ref. \cite{Maruyama}
  gives us a suitable framework to study the dynamical decay process
  of the pentaquark.
  The CMD is a quantum molecular dynamics for constituent quarks,
 in which single quark wave function is parameterized by
  a Gaussian wave pocket in coordinate space and 
  by a color coherent state in the $SU(3)$ color space.
  Time-dependent dynamics of the multi-quark system  is
  then  governed by the Hamiltonian commonly used in the 
    standard constituent quark models \cite{Oka1}. 
  The clusterings and decays of the multi-quarks are easily treated
  by  this approach.  As a first attempt, we will treat the most
  essential part of the decay process (color and spatial coordinates)
   and neglect spin, flavors and antisymmetrization. 
 
 \section{Basic formulations of CMD}
 
We express a total wave function of the system $\Psi$ as a direct product of single-particle quark wave-functions:
\begin{eqnarray}
\Psi &=& \prod_{i=1}^{N} \phi_{i}({\bf r})\chi_{i}, \\ 
\phi_{i}( {\bf r}) &\equiv&   (\pi L^{2})^{-3/4} \exp \left[ - \frac{( {\bf r}- {\bf R}_{i})^{2} }
{ 2L^{2} } - \frac{ i }{\hbar } {\bf P}_{i} \cdot {\bf r} \right] , \\
\chi_i &\equiv&   \left( \begin{array}{@{\,}cccc@{\,}} 
                     \cos \alpha_i \, e^{-i\beta_i} \, \cos \theta_i \\
                     \sin \alpha_i \, e^{\, i\beta_i} \, \cos \theta_i \\
                     \sin \theta_i \, e^{\,i\varphi_i} \\
         \end{array}
      \right)  .
\end{eqnarray}
Here $i$ specifies the quarks and anti-quarks.
 $N$ is the total number of quarks+anti-quarks in the system ($N= 2, 3$ and  5 
  for the kaon, the nucleon, and the  pentaquark, respectively). 
$\phi_i$ is a Gaussian wave packet centered around $ {\bf R}_i$
 with momentum $ {\bf P}_i$ and a fixed width $L$, and
 $\chi_i$ is a coherent state in the color SU(3) space parameterized by 
 four angles, $\alpha_i , \beta_i , \theta_i , \varphi_i$ .

Time evolution of the system is given by solving the equations of motion for 
$\{ {\bf R}_i , {\bf P}_i , \alpha_i , \beta_i , \theta_i , \varphi_i\}$ 
obtained from  the time-dependent variational principle ($\delta L=0$) 
 on the expectation value for the Lagrangian: 
 \begin{eqnarray}
 L &=& \langle \Psi | i\hbar \frac{d}{dt} - \hat{H} |\Psi \rangle \nonumber \\
   &=& \sum_{i} \left[ -\dot{ {\bf P}}_i \cdot {\bf R}_i + \hbar \dot{\beta}_i 
   \cos 2\alpha_i \cos^2 \theta_i - \hbar \dot{\varphi}_i \sin^2\theta_i \right] -H,
 \end{eqnarray}
where $H= \langle \Psi | \hat H |\Psi \rangle$.
The Hamiltonian of the system is given as:
\begin{eqnarray}
\label{eq:hamiltonian}
\hat{H} &=& \sum_{i} \sqrt{m_{i}^2+\hat{{\bf p}}_{i}^{2}} + \frac{1}{2}\sum_{i,j\not=i} \left[ 
-\sum_{a=1}^{8} t_{i}^{a} t_{j}^{a} V_C(\hat{r}_{ij})  + V_M(\hat{r}_{ij}) \right] , \\
V_C(r) &=&  Kr-\frac{\alpha_s}{r}, 
\end{eqnarray} 
where $t^a = \lambda^a /2$ for quarks and $t^a = -\lambda^{\ast a} /2$ for anti-quarks with $\lambda^a$ being Gell-Mann matrices.
$V_C$ consists of one-gluon exchange and confinement terms 
 with an infrared cutoff at $r=3.0\, {\rm fm}$ \cite{Maruyama}.
 Typical values of the parameters in the quark model for
 baryons and mesons read $m_{u,d}=300\, \mathrm{MeV}$, 
$m_{s}=500\, \mathrm{MeV}$ (the constituent-quark mass), 
$\alpha_s = 1.25$ (the QCD fine structure constant), and 
$K=0.75\, \mathrm{GeV/fm}$ (the string tension). 
We take $L_{u,d}$ (size of the quark wave packet) to be $0.49\, \mathrm{fm}$, 
so that the root-mean-square (rms) radius of the quark becomes
 $ 0.6\, \mathrm{fm}$ which  also corresponds to the rms radius of 
 the quark core in the ground state of baryons.
 
 Color-dependent potential $V_C(r)$ alone does not lead to
 the mass difference between $\Theta^+$ and $nK^+$ state. Since
 the 110 MeV mass difference is essential for the pentaquark to
 decay, we introduce $V_M(r)$ to reproduce this mass difference.
   We take  color independent potential with a 
   combination of the attractive scalar-type  and repulsive vector-type 
   Yukawa potentials \cite{Maruyama}.
We assume that $V_M(r)$ does not act on the $\bar{s}$ quark.\footnote{
 In general, the pentaquark can decay to the $nK^+$ state and the $pK^0$ state. 
 However, in the present model, we do not distinguish $u$ and $d$ quarks.
 Therefore,  hereafter we use a notation  ``$NK$" to
  describe  the final decay product.}
  The effective size of the wave packet $L^{\rm eff}$ 
 in the matrix element of $V_{M}$ is chosen to be $0.6\, \mathrm{fm}$ 
 to reproduce the mass splitting between 
 $\Theta^+$ and $NK$.

\subsection{Criterion for color-whiteness}

  To make  a color-white 
  hadron\footnote{Since we use a coherent state in color space,
  color-white does not necessarily imply  color-singlet. 
  The color-singlet part should be projected out from the color-white
  wave function given in the present paper. This is a future problem
   to be done.}
  from arbitrary initial configurations,
  we use a cooling technique in color space \cite{Maruyama}.
  Whether the system of five quarks (four quarks labeled by $i=1\sim4$
   and  one anti-quark labeled by $i=5$)
  become color-white or not is decided by solving the  following criterion; 
\begin{eqnarray}
\sum_{i=1}^{4} \langle \chi_i | \lambda^a | \chi_i \rangle - \langle \chi_5|\lambda^{\ast a} | \chi_5\rangle = 0 
\quad (a=1,\cdots ,8). \label{eq:colorpenta}
\end{eqnarray}
 Criterion of color whiteness for the $q\bar{q}$ and $qqq$ systems are similarly defined.
 The color-white pentaquark satisfying eq.(\ref{eq:colorpenta}) 
  is not necessarily composed of two color-white subclusters.
  This is why dynamical rearrangement of internal colors is required
  for the pentaquark to decay into the $NK$ state.  

\section{Decay of the pentaquark}

\subsection{Color mixing rate ($\alpha$) and effective spatial distance ($D$)} 

The decay of the pentaquark to the $NK$ state is a dynamical process
 in which all the coordinate values (10 for each quark and  50 for the
  five quarks)  defined in the previous section change
  in time. Although CMD simulations trace time-development 
   of all these coordinates, it is convenient to define  some
  particular combinations which can qualitatively
   characterize the decay process.
   $\alpha$ and $D$ defined below are such key parameters.

 $\alpha$  is a measure how well the colors are mixed
 among five quarks. We take notice of an anti-quark in the 
 system and calculate a distance between the anti-quark and 
 one of the other quarks in color-space.
   The minimum value within four possible distances
  is called $\alpha$.  The definition of $\alpha$ in color space can be seen from the 
  following formula:
\begin{eqnarray}
\alpha = \min_{i=1,...,4}\left\{ \sum^{8}_{a=1}\left[ \langle \chi_i | \lambda^a | 
\chi_i \rangle - \langle \chi_5|\lambda^{\ast a} | \chi_5\rangle \right]^2 \right\} . \label{eq:alpha}
\end{eqnarray}
 For $\alpha =0$,  not only the five quarks are color-white as a whole, 
 but also there are color-white $q\bar{q}$ and $qqq$ subclusters.
 As  $\alpha$ increases, the mixture of color in the five quarks
  increases.
   The system cannot be separated into two white clusters any more
  for large $\alpha$. Note that this parameter is defined only in the 
   color space.  Therefore, even if subclusters formed at small $\alpha$
 are color-while, they do not have to be clusters in coordinate space.

 Next, we define an effective distance $D$ which is a measure
  the rate of clustering of the five quarks  in the  coordinate space.
  We consider the anti-quark.
   A quark closest to the anti-quark 
 is supposed to form a subcluster with the anti-quark in the coordinate space.
  Then, we define a center of mass coordinate as ${\bf R}_{q\bar{q}}$.
   Together with the center of mass coordinate of the
   remaining three quarks ${\bf R}_{qqq}$, 
   $D$ is  defined by 
\begin{eqnarray}
D \equiv |{\bf R}_{q\bar{q}}-{\bf R}_{qqq}| . 
\end{eqnarray}
  For small $D$, the five quarks are in  one unit, 
 while for large $D$, they split into two spatially separated
  subclusters. Since  $D$ is defined only in the coordinate space, 
   each subcluster is not necessarily color-white.

 By using these parameters, the pentaquark $\Theta^+$ is characterized
 as a state with large $\alpha$ and small $D$,
  while the $NK$ scattering state is characterized 
   by small $\alpha$ and large $D$.
  Both parameters are time-dependent and their
  initial values at $t=0$ are defined as
 \begin{eqnarray}
 \alpha_{\rm init} \equiv \alpha(t=0), \ \ 
 D_{\rm init} \equiv D(t=0).
 \end{eqnarray} 

We simulate the time development 
of the five quarks from initial conditions with various color
 and spatial configurations ($\alpha_{\rm init}$ and $D_{\rm init}$). Then we
   estimate the lifetime of the system until its decay into 
  the $NK$ state. 

\subsection{Correlation between the color mixing and the life time}

Let us first investigate the correlation between the initial
 color mixing rate and the lifetime.
 We start with 
 initial variables $\{ {\bf R}_i, \alpha_i, \beta_i, \theta_i, \varphi_i\}$  
  randomly chosen under the constraint that the 
  five quarks are  color white as a whole.
  We choose the initial momentum ${\bf P}_i=0$.
 Then, the equation of motion for 50 ($5\times 10$) coordinates are solved
  until the system decays into color-white $q\bar{q}$ ($K$) 
   + color-white $qqq$  ($N$) system.
  The life time is defined when $\alpha$ becomes sufficiently small
   and $D$ becomes sufficiently large so that the system becomes
    the $NK$ state. Actual conditions we use are  
\begin{eqnarray}
\alpha < 0.05 \quad \mathrm{and} \quad D > 3.0\, \mathrm{fm}. \label{eq:collapse_condition}
\end{eqnarray}  

Figure \ref{fig:correlation}(a) shows a relation between 
 $\alpha_{\rm init}$  and the lifetime $T$. 
 The solid point for a given $T$ is obtained by averaging over 
  randomly distributed values of $\alpha_{\rm init}$. 
 The error-bars show the 1$\sigma$ variance of $\alpha_{\rm init}$. 
  This figure clearly shows a positive correlation between
 $T$ and $\alpha_{\rm init}$. The result can be easily 
  understood: it takes more time for the state with large color mixing  
  to rearrange their colors to white and white, which decays into the $NK$ final state.

\subsection{Correlation between spatial clustering and the life time}

\begin{figure}
 \begin{center}
\includegraphics[width=15cm]{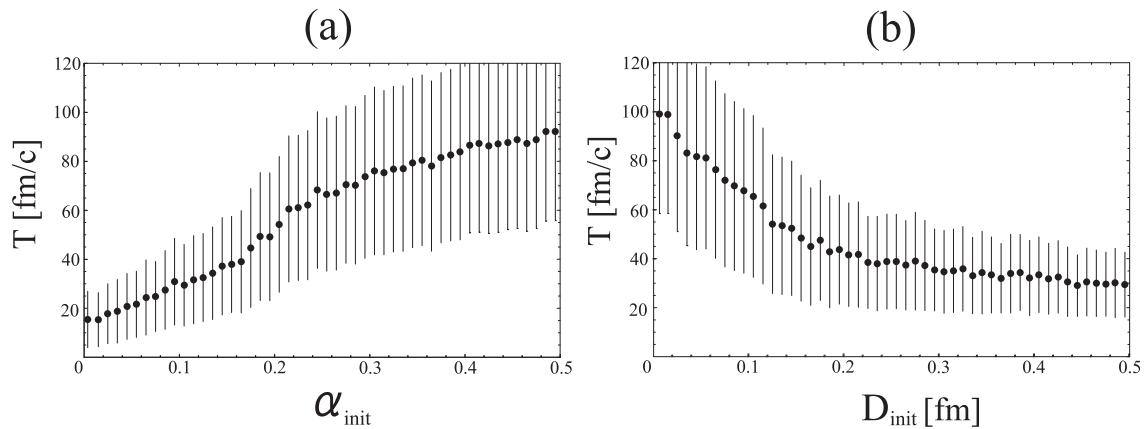}
\caption{(a) Relation between the initial color mixing rate
  $\alpha_{\rm init}$ 
and the lifetime $T$. (b) Relation between the initial spatial distance
 $D_{\rm init}$ and the lifetime $T$.}
\label{fig:correlation}
 \end{center}
\end{figure}

Next we investigate the relation between the initial
 spatial distance $D_{\rm init}$ and the lifetime $T$.
 Figure \ref{fig:correlation}(b) shows the relation between
 $D_{\rm init}$ and the lifetime $T$.
The samples are the same as in the case of Sec.3.2.   
The  solid point for a given $T$ is obtained by averaging over 
 randomly distributed  values of $D_{\rm init}$. 
 The error-bars show the 1$\sigma$ variance of the
  $D_{\rm init}$.
  The figure shows a negative correlation between
 $T$ and $D_{\rm init}$. The result can be easily 
  understood: it takes more time for the state with small
   spatial distance between subclusters 
  to rearrange their coordinates to the $NK$ final state.

If we choose   $\alpha_{\rm init}=0.5$ and 
 $D_{\rm init}=0$  as a typical parameter set for the 
  pentaquark state $\Theta^+$, we have the life time 
  (decay width) of about 100 fm/c (2 MeV). It is
   too hasty at the moment to compare this number with the experimental
    upper bound of 10 MeV, but the result is
     suggestive.

\subsection{Potential surface as a function of  $\alpha$ and $D$}

To study the effect of color mixing rate $\alpha$ and the effective spatial
 distance $D$ on the dynamical decay process in more  details,
 we calculate an effective  potential $V$  between the $q\bar{q}$ and $qqq$    subclusters
  as functions of $\alpha$ and $D$.
 
\begin{figure}
 \begin{center}
\includegraphics[width=15cm]{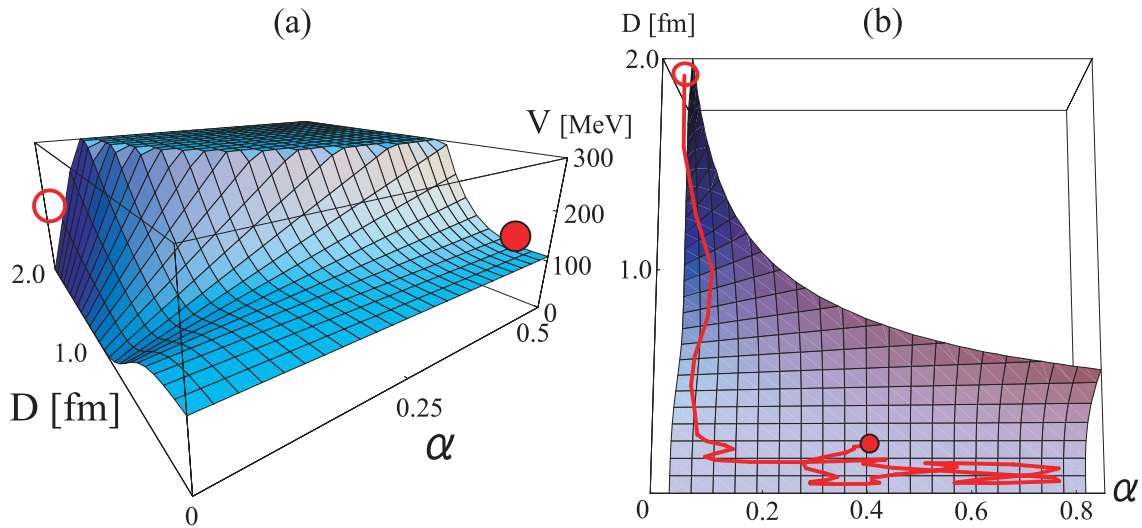}
\caption{(a) An effective potential $V$ on the $\alpha-D$ plain.
 (b) Actual motion of the pentaquark decaying into the $NK$ state is
  superimposed on the top view of the effective potential surface.}
\label{fig:surface}
 \end{center}
\end{figure}

  Shown in Fig. \ref{fig:surface}(a) is the result of such calculation
 using the potential part of the Hamiltonian in eq.(\ref{eq:hamiltonian}). 
  In practice, $V(\alpha,D)$ is calculated as follows: 
 first, we prepare a $q\bar{q}$ state with quark and anti-quark 
  sitting on top of each other in coordinate space, and  
  $qqq$ is prepared in the same way. Then, we calculate $\alpha$, $D$
   and the expectation value of the potential energy of this state.
   Then, this process is repeated to cover the variety of points in 
   two-dimensional $\alpha-D$ plane.
 With this procedure,  $V$ is adjusted to be zero for the 
 $NK$ final state  at large $D$ and small $\alpha$.
 Also $V$=110 MeV for small $D$ and large $\alpha$ corresponds to
  the ideal excitation energy of the pentaquark state $\Theta^+$.

 In general, $V$ increases as $D$ increases due to the 
  effect of color confinement potential.  Namely, it takes more and more
   energy when one tries to separate the five quarks into
    color non-white subclusters in the coordinate space.
 The only exception is $\alpha=0$ where
  the potential energy decreases as $D$ 
 increases. This is  because there is no resistance from the 
 confinement force in this case. 
 Note also that the potential surface is flat in the $\alpha$ direction
  for small $D$. 
   (One can prove that $V$ is exactly $\alpha$-independent for $D=0$).
  This implies that the color-dependent potential become important
   only when quarks are separated in space.
   
 Now, suppose that the pentaquark state $\Theta^+$ is located
  at small $D$ and large $\alpha$ region as indicated by the 
   solid circle in
   Fig. \ref{fig:surface}(a).
  For this state to decay into the $NK$ state indicated
   by the open circle  in
   Fig. \ref{fig:surface}(a), it has to rearrange color 
   to find a narrow channel near $\alpha=0$.  This takes a long time
    since the potential surface is flat in $\alpha$ direction and thus
    the system goes back and forth before reaching the channel. 
  Once it reaches the region around $\alpha=0$, 
  it quickly decays into the $NK$ state along the narrow channel.
  Namely, the flat 
   potential along the $\alpha$-axis near $D=0$ and
    the narrow channel along the $D$-axis
   near $\alpha=0$ are  two essential sources of the 
    long life time of the pentaquark.

Figure \ref{fig:surface}(b) shows the actual path in the
 simulation of  the pentaquark state decaying into the $NK$ state.
The initial conditions are taken to be $\alpha_{\rm init}=0.4$ 
and $D_{\rm init}=0.25\, \mathrm{fm}$, 
and it takes  $25\, \mathrm{fm/c}$ to decay. 
 Wandering in color space of the pentaquark before the decay 
  can be seen explicitly
  from this figure.

\section{Summary}

  In order to study the narrow decay width of the pentaquark
  suggested by experiments,
  we have carried out color molecular dynamics (CMD) simulation
  for five quarks.
  In CMD, the spatial and color parts of the wave function for each quark is expressed as the
  Gaussian wave packet and $SU(3)$ coherent state respectively. An advantage
  of this approach is that one can trace the rearrangement process of
  spatial and color coordinates as a function of time
  during the decay.

  To characterize the essential part of the decay process, we have introduced 
  two key parameters, the color mixing ratio $\alpha$ and the 
  effective spatial distance $D$.
  The results of the simulation show that there is a positive (negative)
  correlation between $\alpha(t=0)$ ($D(t=0)$) and the life time $T$
  of the five-quark state.  $T$ can reach to even 100 fm/c if $\alpha(t=0)$ 
  is enough large and $D(t=0)$ is enough small. Narrow channel in the effective potential surface
   $V(\alpha,D)$ is found to be the physical origin to cause the 
   long life time.  The pentaquark wanders around the potential surface.
 
  In the present paper, we have not considered the spin and flavor
  of the quarks, and hence the spin and flavor dependent interactions
  in the Hamiltonian.  The antisymmetrization of the quark wave functions
  are also neglected for simplicity.   
    Detailed study of the decay process taking into account these
  ingredients is an urgent problem to be 
  examined.\footnote{Spin, flavor and antisymmetrization
   were recently taken into account in the
    antisymmetric molecular dynamics (AMD)
  study of a static pentaquark state in ref.\cite{enyo}. However,
    the decay process is considered only
     in a static and approximate way in that work.}

  In this paper, we have simply assumed that the pentaquark $\Theta^+$
  corresponds to a state with a large $\alpha$ and small $D$.
  Whether one can indeed reach such parameter region by the 
   process $N+K \rightarrow \Theta^+ \rightarrow N+K$ can be
   studied by the scattering simulations in CMD, which
    will be reported elsewhere \cite{maezawa-2}.
  Applications of the present study to the decay processes of
   other possible narrow resonances such as charmed pentaquark
    are also the interesting problems to be 
    investigated.

\section*{Acknowledgments}
We thank S. Sasaki, A. Hosaka, Y. Akimura and  S. Chiba
for fruitful discussions on the physics of pentaquarks. 
This work was partially supported by the Grants-in-Aid of the
Japanese Ministry of Education, Culture, Sports, Science, and Technology
(No.~15540254).

\vfill\eject
\end{document}